\documentclass[5p,final,times,sort&compress]{elsarticle}
\pdfoutput=1
\usepackage{algorithmic}
\usepackage{algorithm}
\usepackage{amsmath}
\usepackage{amssymb}
\usepackage{graphicx}
\usepackage{sistyle}
\usepackage{textcomp}
\usepackage{url}
\usepackage{mdwtab}
\usepackage{dcolumn} 

\SIproductsign{\!\times\!}

\usepackage{color}
\usepackage[normalem]{ulem}

\renewcommand\epsilon{\varepsilon}
\renewcommand\phi{\varphi}
\renewcommand\theta{\vartheta}
\renewcommand\vec[1]{\boldsymbol{\mathrm{#1}}}

\newcommand\diff{\mathrm{d}}
\newcommand\dotprod{\!\boldsymbol{\cdot}\!}

\newcommand\expect[1]{\left\langle\vphantom{\big(}#1\right\rangle}
\newcommand\dexpect[1]{\left\langle\left\langle\vphantom{\big(}#1\right\rangle\right\rangle}

\newcommand\eq[1]{Eq.~\eqref{eq:#1}}
\newcommand\fig[1]{Fig.~\ref{fig:#1}}

\newcommand\kB{k_\text{B}}
\newcolumntype{d}[1]{D{.}{.}{#1}}

\def\naive{na\"{\i}ve}

\makeatletter
\newcommand\dash{\penalty\@M-\hskip\z@skip}
\makeatother

\begin{document}

\newcolumntype{C}[1]{>{\centering}m{#1}}

\begin{frontmatter}
  \journal{Computer Physics Communications}

  \title{Highly accelerated simulations of glassy dynamics using GPUs:
caveats on limited floating-point precision}

  \author[asc,dlr]{Peter H. Colberg\fnref{cptg}}
  \ead{peter.colberg@utoronto.ca}

  \author[asc,oxf]{Felix H{\"o}f{}ling\fnref{mf}}
  \ead{hoefling@mf.mpg.de}

  \address[asc]{Arnold Sommerfeld Center for Theoretical Physics
  and Center for NanoScience (CeNS), Fakult{\"a}t f{\"u}r Physik,
  Ludwig-Maximilians-Universit{\"a}t M{\"u}nchen,
  Theresienstra{\ss}e 37, 80333 M{\"u}nchen, Germany}

  \address[dlr]{Institut f{\"u}r Materialphysik im Weltraum,
  Deutsches Zentrum f{\"u}r Luft- und Raumfahrt (DLR),
  Linder H{\"o}he, 51147 K{\"o}ln, Germany}

  \address[oxf]{Rudolf Peierls Centre for Theoretical Physics, 1 Keble Road,
  Oxford OX1 3NP, England, United Kingdom}

  \fntext[cptg]{\emph{Present address:} Chemical Physics Theory Group,
  Department of Chemistry, University of Toronto, Toronto,
  Ontario M5S 3H6, Canada}

  \fntext[mf]{\emph{Present address:} Max-Planck-Institut f\"ur Metallforschung,
  Hei\-sen\-berg\-stra{\ss}e 3, 70569
  Stuttgart and Institut f\"ur Theore\-tische und Angewandte Physik, Universit\"at
  Stuttgart, Pfaffenwaldring 57, 70569 Stuttgart, Germany}


\begin{abstract}
    Modern graphics processing units (GPUs) provide impressive computing resources,
    which can be accessed conveniently through the CUDA programming interface.
    We describe how GPUs can be used to considerably speed up molecular
    dynamics (MD) simulations for system sizes ranging up to about 1 million particles.
    Particular emphasis is put on the numerical long-time stability in terms of energy
    and momentum conservation, and caveats on limited floating-point precision
    are issued.
    Strict energy conservation over \num{E8} MD steps is obtained by
    double-single emulation of the floating-point arithmetic in
    accuracy-critical parts of the algorithm.
    For the slow dynamics of a supercooled binary Lennard--Jones mixture,
    we demonstrate that the use of single-floating point precision may result
    in quantitatively and even physically wrong results.
    For simulations of a Lennard--Jones fluid, the described implementation shows
    speedup factors of up to 80 compared to a serial implementation for the CPU,
    and a single GPU was found to compare with a parallelised MD simulation
    using 64 distributed cores.
\end{abstract}
\end{frontmatter}

\section{Introduction}

In recent years, the computational power of graphics processing units (GPUs) has
increased rapidly: the theoretical peak performance for single precision
floating-point operations on an amateur's GPU\footnote{NVIDIA GeForce GTX 280}
reaches nearly \SI{1}{Tflop/s}.
Compared to a single core of a conventional processor,
this gives rise to an expected performance jump of one or two orders of
magnitude for many demanding computational problems, fueling the desire to
exploit graphics processors for scientific applications.
While conventional high-performance computing (HPC) depends on expensive central
computing clusters, shared amongst many researchers, GPU computing
makes local clusters acting as small HPC facilities conceivable for large-scale
simulations at a fraction of the cost.

A modern GPU works as a streaming processor implementing the
single\dash{}instruction\dash{}multiple\dash{}threads (SIMT) model.
One device contains several hundred scalar processor cores
executing a single instruction or a small set of divergent instructions
in parallel in thousands of threads on a large data set.
Parallel, coalesced access to the onboard device memory via a memory interface
of up to 512 bits provides a remarkably high memory bandwidth.

Molecular dynamics (MD) simulations---a widespread tool for particle-based
simulations in biological physics, chemistry, and material science---are
ideally applicable to the GPU due to their inherent parallelism.
The release of NVIDIA's \emph{compute unified device architecture} (CUDA) as a
convenient means of GPU programming has triggered a lot of activity in
exploiting GPUs for scientific applications.
Several MD implementations using GPUs have demonstrated significant speedups,
with performance measurements based on relatively short test
runs~\cite{Anderson2008, Meel2008, Liu2008, Stone2007, Yang2007}.
A critical review of current attempts to exploit GPUs in MD simulations
may be found in Ref.~\citenum{Xu:2010}; in essence, published results are not
yet available, showing the ``nascent nature'' of the field.
As a notable exception, complex MD simulations targeting at protein folding
were accelerated by GPUs allowing for the trajectories to reach into the
millisecond regime~\cite{Voelz:2010, Harvey:2009}.
In the realm of physics, we are only aware of a Monte-Carlo study of the critical
behaviour of the Ising model that was performed on the GPU, reporting speedups
between 35 and 60 for the mostly integer arithmetic-based algorithm~\cite{Preis2009};
a multi-GPU approach to this problem shows a promising scalability~\cite{Block2010}.

In this article, we describe MD simulations that are fully implemented on the GPU
and that faithfully reproduce the glassy dynamics of a supercooled binary mixture.
If a glass-forming liquid is cooled or compressed, the structural relaxation
critically slows down by several orders of magnitude and a rapidly growing
time scale emerges close to the glass transition line~\cite{Goetze:MCT}.
Small changes in the temperature may already have drastic effects on the
dynamics, and a numerical study thus requires excellent long-time stability
with respect to energy conservation over long simulation runs
of \SI{e8} MD integration steps and more.
The single precision floating-point arithmetic provided by recent GPUs turns out
to be a major obstacle to this goal. But once this limitation is overcome,
general-purpose computing on GPUs provides a useful tool to address current
questions of the glass transition with minimal allocation of hardware and
reasonable computing time.

The substantial enhancement of computing resources
due to GPU computing facilitates the investigation of very large systems,
which is desirable for studies of phenomena associated with a
divergent length scale.
Only recently, evidence for a divergent static correlation length in
supercooled liquids was found in large-scale simulations of
up to 64,000 particles~\cite{Mosayebi2010}
and 80,000 particles~\cite{Flenner2010}, respectively.

The article is organised as follows.
We describe a high-precision implementation for the GPU in
Section~\ref{sec:implementation}, followed by performance benchmarks
in Section~\ref{sec:performance}.
A detailed analysis of the long-time stability regarding momentum and energy
conservation is given in Section~\ref{sec:stability}.
Section~\ref{sec:glassy} demonstrates the impact of numerical accuracy
on the simulation results for the glassy dynamics of a binary mixture of soft spheres.
A conclusion is given in Section~\ref{sec:conclusion}.

\section{Implementation for the GPU}
\label{sec:implementation}

\subsection{Architecture of the GPU hardware}

An MD implementation for the GPU needs to be adapted to its vastly different
architecture.
A modern GPU consists of hundreds of scalar cores, which map to execution
threads in the CUDA architecture and require fine-grained parallelisation of the
algorithm.
The scalar cores are divided into multiprocessors---units of 8 processors for
the NVIDIA G200 series GPUs---capable of executing an atomic warp of 32 threads
in four clock cycles. Each multiprocessor is equipped with a total of 16,384
32-bit registers, and a fast, tiny shared memory of 16\,kB;
all multiprocessors access a large global device memory of 1 to 4\,GB, which is
two orders of magnitude slower than local registers.
To hide latencies when accessing the global device memory, a scheduler
concurrently executes multiple warps on a multiprocessor.
The equivalent of a multiprocessor in the CUDA architecture is a block of up to
512 threads, and the only means of communication and synchronisation is within
a block.
For the execution of complex algorithms, the number of threads per block has to
be lowered due to the limited number of registers available per multiprocessor.
Global device memory enforces a strict memory access pattern, where threads have
to address memory in a linear \emph{coalesced} order.
In contrast to shared memory access, random read and write access to global device
memory entails a performance penalty of an order of magnitude.
This impairs GPU acceleration of many common computational primitives such as
sorting algorithms.
As a partial remedy a texture cache of up to 8\,kB per multiprocessor, which
stems from the graphics heritage of the GPU, enables read-only random access to
a small window of global device memory.
A global cache of 64\,kB of constant memory provides access to constant data at
speeds comparable to register access.

\subsection{The MD integration step}

A soft-sphere molecular dynamics (MD) simulation solves Newton's equations in
discretised time for $N$ classical particles interacting via a $C^2$-continuous,
short-ranged potential.
Every MD step, the force on each particle exerted by all interacting
particles is calculated, and the particles are propagated by a symplectic
integrator, the velocity-Verlet algorithm~\cite{Rapaport2004}.
A \naive{} implementation based on a doubly nested force loop would yield an
algorithmic complexity of $O(N^2)$.
For short-ranged interparticle forces however, a linear scaling of the
performance with the number of particles can be achieved with Verlet neighbour
lists.
For each particle, a list of particles located within a radius $r_c$, the
cutoff radius of the potential, is kept in
memory; the force algorithm then considers only particles in the neighbour list.
A small ``skin'' of a fraction of $r_c$ is added to the neighbour sphere
to reduce the necessity of rebuilding the neighbour lists to every 10 to 100 steps.
Particle binning---the decomposition of the system into spatial cells---avoids a
doubly nested loop in the neighbour list algorithm by limiting the search for
neighbours of a particle to its own and the adjacent cells.
Our implementation for the GPU combines and extends concepts described in detail
in Refs.\ \citealp{Anderson2008} and~\citealp{Meel2008}.

Parallelisation of the velocity-Verlet algorithm for the GPU is straightforward
and naturally respects coalesced memory access:
each CUDA thread is assigned a single particle with the task of updating the
velocities and coordinates.
The implemented force and neighbour list algorithms resemble the ones proposed
in Ref.~\citealp{Anderson2008}.
Specifically in the force algorithm, each thread reads the indices of a
particle's neighbours in a coalesced manner, and coordinates are then fetched
using the texture cache, mitigating performance penalties due to random memory
access.

For particle binning, we have implemented a cell list algorithm based on a parallel
radix sort~\cite{Knuth1998, Zagha1991}.
Each particle is assigned a one-dimensional integer cell index.
An array containing the particle numbers is sorted according to the cell
indices, as proposed in the Particles example of the CUDA SDK~\cite{CUDA_SDK}.
A further pass determines the start index of each cell, followed by the assembly
of fixed-size cell lists for the subsequent neighbour list algorithm.
With this method we avoid the bottleneck of a host to device memory copy which
is needed with the CPU-based cell list algorithm of Ref.~\citealp{Anderson2008}.
The radix sort algorithm performs worse on the GPU than on the CPU for
small systems of considerably less than \num{E4} particles, and further, particle
binning is only necessary every 10 to 100 steps. Thus, the overall performance of the
MD step is best for systems of \num{E5} and more particles.

\begin{figure*}
  \centering{\includegraphics{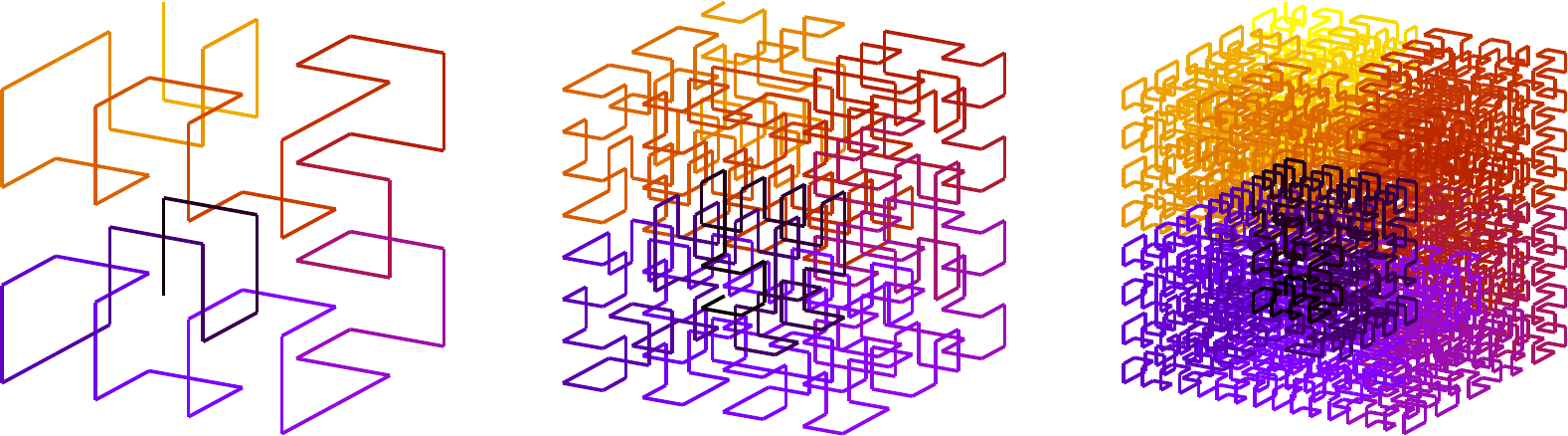}}
  \caption{Hilbert's space-filling curve in three dimensions after 1, 2 and 3 recursions.
  It provides the mapping between three-dimensional space and one-dimensional
  memory, which ensures texture locality of neighbouring particles. Particles at vertices
  with similar colour will be close in memory after the sort.}
  \label{fig:hilbert}
\end{figure*}

The neighbour list algorithm finally reads the particle indices from the
fixed-size cell lists to gather coordinates and velocities of particles in
neighbouring cells via texture fetches, temporarily stores them in shared
memory, and builds a fixed-size neighbour list~\cite{Anderson2008}.

A parallel reduction scheme\footnote{See example ``Reduction'' in
Ref.~\citealp{CUDA_SDK}} on the
GPU is used to calculate properties which are needed every MD step such as
the maximum absolute particle displacement and the potential energy sum as well
as less frequently evaluated properties such as temperature, centre-of-mass
velocity, and the virial tensor.

In the force algorithm, we make efficient use of the texture cache by
periodically sorting the particles in global device memory according to a
three-dimensional space-filling Hilbert curve~\cite{Sagan1993, Anderson2008}.
Such a curve provides a mapping between 3D space and 1D memory
to optimally conserve spatial locality.
In our implementation, the mapping is recursively generated on the GPU using 8 vertex
rules~\cite{Wang2005} as depicted in \fig{hilbert}, based on a lattice
spacing of $\sigma$ with a maximum recursion depth of~10.
The subsequent particle sort employs the radix sort algorithm~\cite{Knuth1998,
Zagha1991} to generate a permutation array; then read access to the texture
cache and coalesced write access to global memory are used to efficiently
permute particle coordinates and velocities.

\subsection{Random number generator}

A modified Andersen thermostat for pre-equilibration cooling is realised by
assigning Boltzmann-distributed velocities to all particles every
$(\mu\,\delta t)^{-1}$ steps according to a fixed heat bath collision
rate $\mu$.
Furthermore, the assigned velocities are rescaled by means of parallel reduction
to exactly match the temperature of the heat bath.
We have implemented the parallel \textsf{rand48} random number generator, which
may be trivially parallelised by leap-frogging within the sequence~\cite{Meel2008}.
The linear recurrence $x_{t+T} = A_T x_{t} + C_T \mod 2^{48}$ uses a leapfrog
multiplier $A_T = a^T \mod 2^{48}$ and leapfrog addend
$C_T = c \sum_{n = 0}^{T-1} a^n \mod 2^{48}$ to jump within the sequence
depending on the total number of threads~$T$.
As an improvement compared to Ref.~\citealp{Meel2008}, we seed the parallel
generator using the parallel prefix sum algorithm\footnote{See example ``Scan''
in Ref.~\citealp{CUDA_SDK}}, which yields the initial state $x_t$ for each
thread~$t$ by binary-tree summation and multiplication.
Thus, initialisation times of many seconds are avoided for large systems.

\subsection{Double-single precision floating-point arithmetic}

It will be demonstrated below that numerical long-time stability requires double
precision arithmetic in critical parts of the MD step.
Currently prevalent GPUs of the NVIDIA GT200 series feature only
1~double precision floating-point unit per every 8 single precision units, which
prohibits the use of native double precision in performance-critical algorithms.

The limited native precision is overcome by algorithms
which implement multi-precision arithmetic using two native floating-point
registers~\cite{Dekker1971, Knuth1997}.
On hardware supporting the IEEE 754-1985 floating-point
standard~\cite{IEEE754-1985}, proofs of numerical exactness have been
given for multi-precision addition and multiplication~\cite{Lauter2005}.
For the GPU, which does not fully comply with IEEE 754\dash{}1985 in terms of
rounding and division, these proofs have been adapted for double-single
precision arithmetic using two native single precision
floats~\cite{Gracca2006}.
The double-single precision algorithms for addition and multiplication with
NVIDIA GPUs yield an effective precision of 44~bits.

\begin{algorithm}[tb]
  \caption{Algorithm for the addition of two floating-point numbers in
  double-single precision, based on the DSFUN90 package~\cite{Bailey2005}. The
  subscripts 0 and 1 refer to the high and low word of a double-single
  float, respectively.}
  \label{alg:dsfun90_addition}
  \begin{algorithmic}
    \STATE \COMMENT{Compute $a + b$ using Knuth's trick.}
    \STATE $t_0 \gets a_0 + b_0$
    \STATE $e \gets t_0 - a_0$
    \STATE $t_1 \gets ((b_0 - e) + (a_0 - (t_0 - e))) + a_1 + b_1$
    \STATE
    \STATE \COMMENT{The result is $t_0 + t_1$, after normalisation.}
    \STATE $c_0 \gets e \gets t_0 + t_1$
    \STATE $c_1 \gets t_1 - (e - t_0)$
  \end{algorithmic}
\end{algorithm}

For IEEE 754\dash{}1985 compatible hardware, the DSFUN90 package for
Fortran~\cite{Bailey2005} contains implementations of addition,
subtraction, multiplication, division, and square root in double-single
precision. We have ported these implementations from Fortran to CUDA, extending the
functionality provided by the Mandelbrot example of the CUDA SDK~\cite{CUDA_SDK}.
As an example the addition algorithm in double-single precision is displayed in
Table~\ref{alg:dsfun90_addition}.
To implement double-single multiplication on the GPU, care has to be taken to
avoid fused multiply-add operations, as the GPU does not round the result of
the comprised multiplication, in violation of the floating-point standard.
As a remedy, CUDA provides the intrinsic operations \textsf{\_\_fmul\_rn} and
\textsf{\_\_fadd\_rn}.

In double-single precision, we have implemented the update of velocities and
coordinates in the velocity-Verlet step as well as the summing over force
contributions from neighbouring particles.
These may be subject to accumulated summing errors depending on the
particular time step and potential, respectively.
The evaluation of the force contributions itself, which accounts for a
large fraction of the computational cost within the MD step, remains in single
precision.

\subsection{Evaluation of time-correlation functions}

The dynamic properties of a molecular system are often quantified in terms of
time-correlation functions~\cite{Hansen:SimpleLiquids}.
For observables $A$ and $B$, one defines their correlation at different times
$t$ and $s$ as
\begin{align}
C_{AB}(t,s) & =\expect{A(t) B^*(s)} \notag \\
& =\lim_{T\to\infty}\frac{1}{T}\int_0^T\!\diff \tau A(t+\tau)\,B^*(s+\tau).
\label{eq:TCF}
\end{align}
In equilibrium, time-correlation functions are stationary and depend only on the
difference $t-s$, thus $C_{AB}(t,s)=C_{AB}(t-s)$.
The most important class of time-correlation functions is comprised by the
\emph{auto}-correlation functions $C_{AA}(t)$.
If the system is ergodic, the average may be evaluated alternatively as an
ensemble average over initial conditions, i.e., over independent simulation
runs.
For tagged-particle observables, the statistical error is decreased additionally
by averaging over all particles of the same species.

\begin{figure}
\centering{\includegraphics{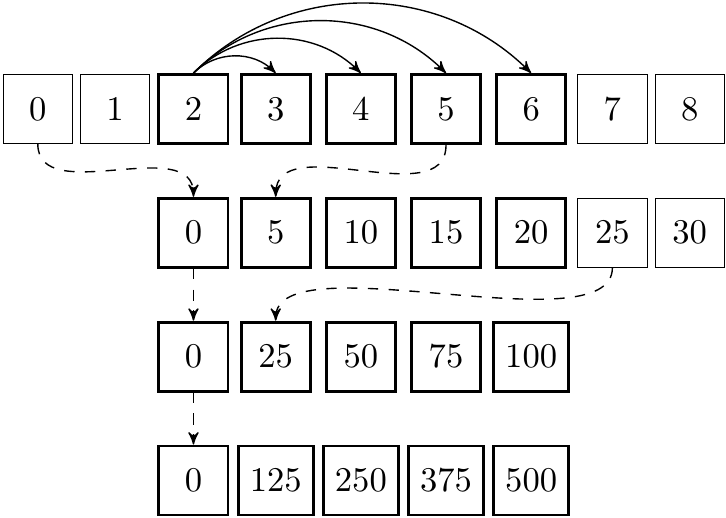}}
\caption{Illustration of the blocking scheme for $k=4$ blocks of size $\ell=5$.
Each square represents the system state at a given time;
the numbers refer to the finest time resolution $\Delta t$.
Dashed arrows indicate the filling of the higher block levels with lower resolution.
Solid lines indicate the correlations calculated between the most left entry and
all other entries of the same block.
The thin squares to the left have already been processed and were discarded,
while those to the right still have to be inserted into the blocks.}
\label{fig:blocking_scheme}
\end{figure}

We have developed a blocking scheme which allows the online evaluation of
time-correlation functions during the production run.
The scheme was inspired by the ``order-$n$ algorithm'' of Ref.~\citealp{Frenkel:MD}.
Complex relaxation processes often comprise several time scales and are usually
discussed on a logarithmic time axis.
The blocking scheme yields the correlation functions for fast and slow processes
simultaneously and provides a time grid already suitable for a logarithmic
representation.
Introducing some short-time resolution $\Delta t$, we assume that the state of
the system is stored at times $m \Delta t$ for $m=0,1,2,\dots$, from which the
observables $A_m=A(m\Delta t)$ are calculated.
For a long simulation run over a time span of $T=M\Delta t$,
we approximate the integral in \eq{TCF} by a sum,
\begin{equation}
C_{AA}^{(m)}:=C_{AA}(m\Delta t)\approx
\frac{1}{M'}\sum_{j=0}^{M'-1} A_{m+j}  A^*_j
\end{equation}
where $M'=M-m$.
The evaluation of $C_{AA}^{(m)}$ for all $m$ would require handling a huge
number of $M$ copies of the system state at different times (or of the derived
observables at least) and would involve $O(M^2)$ floating-point operations.
Instead, we arrange the time grid in $k$ blocks of size $\ell$, where the time
resolution between subsequent blocks is increased by a factor of~$\ell$, see
\fig{blocking_scheme}.
Within block~$n$, correlations are calculated for time lags  $\Delta t_n, \dots,
(\ell-1)\Delta t_n$ only, where $\Delta t_n=\ell^n \Delta t$.
The blocks are continuously filled during the simulation.
Whenever a block is full, the first entry is correlated with all other entries
and is then discarded.
Thus, each block contains a section of the trajectory, moving forward as the
simulation progresses.
The memory requirement to handle a trajectory of length $\ell^k \Delta t$ is
merely $k \times \ell$ snapshots of the system state.

\section{Performance measurements}
\label{sec:performance}

A central argument for the use of GPUs in high-performance computing is
their high theoretical peak performance compared to that of a single core of a
conventional Opteron or Xeon CPU. We did extensive performance measurements of
our MD implementation, which we compare first to our own serial reference
implementation for the host processor.
The comparison between a massively parallel implementation and a serial one
is somewhat unfair, in particular in view of the multi-core nature of current CPUs.
Thus, we secondly provide a comparison with the LAMMPS
package~\cite{Plimpton1995}, being one of the established, parallelised MD packages
widely used in the physics community.
The test includes the parallel use of multi-core CPUs within a single compute node
as well as distributed computing over several nodes, which is more realistic for production
use and allows for a more reasonable comparison between a conventional cluster and
a single GPU; a similar approach was taken recently by Harvey \emph{et al.}~\cite{Harvey:2009}.
LAMMPS offers some GPU acceleration as well which we have not used by purpose
and which we explicitly do not refer to.

\begin{figure*}
  \centering{\includegraphics{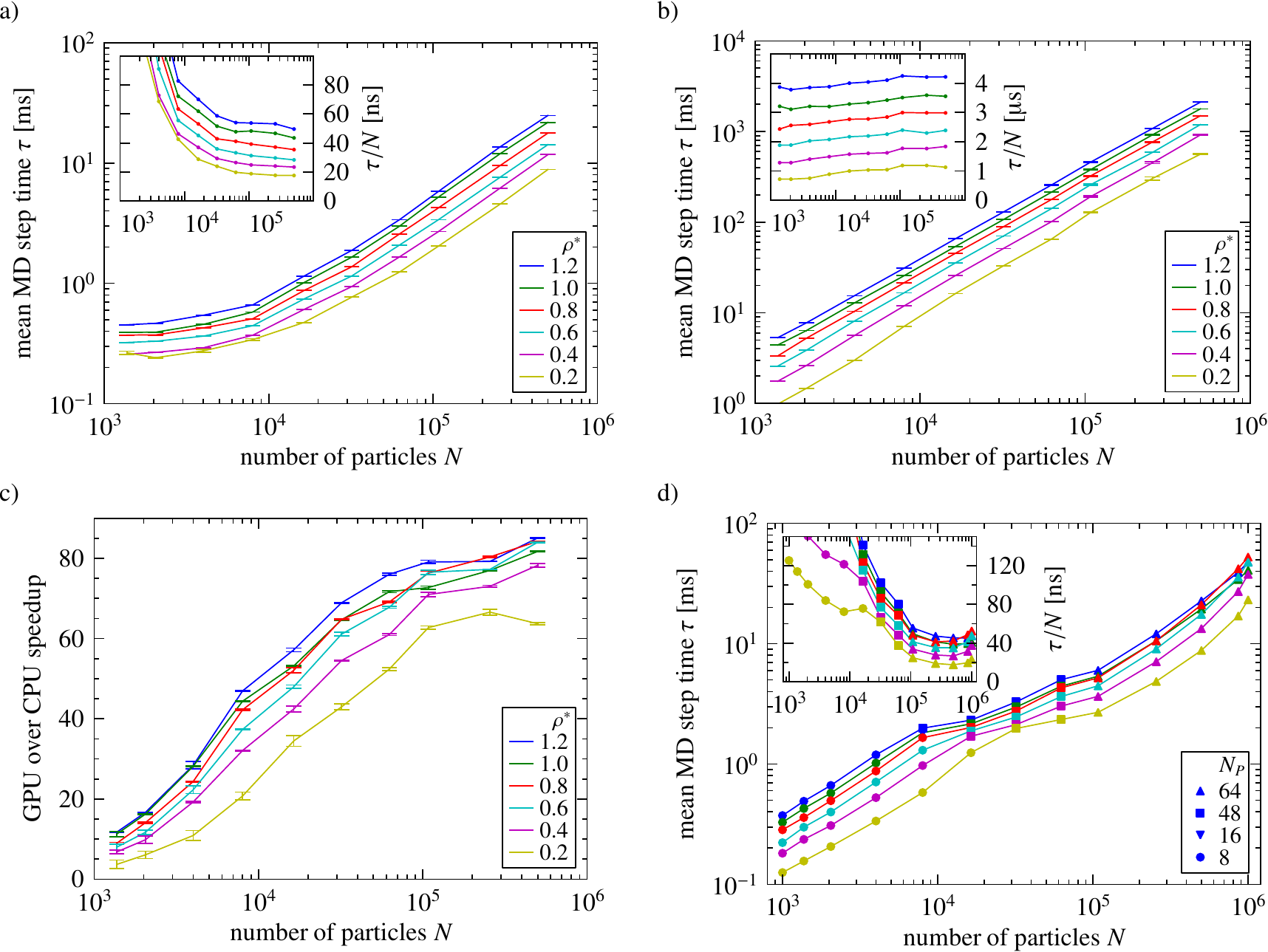}}
  \caption{Results of the performance benchmarks for a Lennard--Jones liquid of various
  densities as function of the number of particles $N$;
  temperature was kept fixed at $T^*=1.5$.
  Large panels show the total time $\tau$ per MD step,
  while insets test the expected linear scaling and display $\tau/N$.~
  Panel~a: average time required for a single MD step on the GPU (single precision);~
  panel~b: time per MD step on a serial host implementation; note the different scale of the
  $y$-axis compared to panels a and d;~
  panel~c: relative GPU performance versus a single CPU core;~
  panel~d: time per MD step using the parallelised LAMMPS package for jobs of various
  process numbers~$N_p$. For each density and particle number, the fastest result is shown,
  where symbols indicate the selected~$N_p$.}
  \label{fig:performance}
\end{figure*}

Our measurements were done on Lennard--Jones liquids of varying size and density
with the conventional 12--6 interaction potential,
$V_\text{LJ}(r;\sigma,\epsilon) = 4\epsilon[(r/\sigma)^{-12}-(r/\sigma)^{-6}]$,
cut off at $r_c=2.5\sigma$;
in LAMMPS we used the potential implementation termed \textsf{lj/cut}.
Units of length, mass, time, and energy are $\sigma$, $m$,
$\sqrt{m\sigma^2/\epsilon}$, and $\epsilon$ as usual, and dimensionless
quantities are indicated by an asterisk.
We used a step size of $\delta t^*= 0.001$ and a neighbour list skin of width
$0.5\sigma$ for the GPU and $0.3\sigma$ for the serial CPU and parallel LAMMPS
benchmarks.
To generate a homogeneous system, an initial fcc lattice was equilibrated in the
NVT ensemble at $T^* = \kB T/\epsilon = 1.5$ for \SI{e4}{} steps.
Then, the wall time spent on the MD step was measured for the next \SI{e4}{}
steps.
For the GPU and parallel CPU benchmarks, the results were averaged over
10~independent realisations.
For the LAMMPS benchmarks, only a single system was generated and the results
were averaged over two consecutive measurement runs of \SI{e4}{} steps each.
The investigated system sizes ranged from $N=1{,}372$ to 864,000 particles and
the densities from $\varrho^*=0.2$ to $1.2$.

The GPU benchmarks were run on GPUs of type NVIDIA GeForce
GTX 280, which contains 240 scalar processor cores clocked at \SI{1.3}{GHz},
with GDDR3 memory at \SI{1.1}{GHz} providing a device memory bandwidth of 140
GB/s ($=2\times\SI{1.1}{GHz}\times\SI{512}{bits}$).
Its theoretical peak single precision floating-point performance based on one
fused multiply-add operation and one concurrent multiply operation per cycle is
\SI{933}{Gflop/s} ($=240\times\SI{3}{flop}\times\SI{1296}{MHz}$).
For comparison with our host reference implementation, we used an AMD Opteron
2216~HE processor at \SI{2.4}{GHz} with PC2-4200 CL4 dual-channel memory and a
theoretical memory bandwidth of 17 GB/s ($=2\times2\times\SI{533}{MHz}\times\SI{64}{bits}$).
The theoretical peak performance on the assumption of a single SSE instruction
execution with four concurrent single precision floating-point operations per
cycle is \SI{9.6}{Gflop/s} per CPU core.
This roughly yields a factor of 100 for the theoretical limit on the speedup of
a compute-bound algorithm, and a factor of 8 for an algorithm limited by memory
bandwidth.
The estimate does not take into account the memory latency on the GPU, which is 400\,--\,600
clock cycles for a thread accessing global memory \cite{CUDA_Guide}, and may
more substantially constrain the overall performance of an algorithm than floating-point
performance or memory bandwidth.
Further, recent CPUs contain several cores, and typical nodes in a
computing centre are found to be 4- to 16-way SMP machines, and one has to put these
numbers into perspective for a parallelised implementation making use of all
available cores simultaneously.

The GPU benchmarks were performed with HAL's MD package \cite[commit e611734]{Colberg2009}
using single floating-point precision on the GPU for comparison with other work,
i.e., without the implementation of double-single precision described above.
The GPU-specific CUDA code was compiled with NVIDIA's CUDA compiler (version~2.2)
targeting compute capability~1.0.
The host-specific code was compiled for the x86\_64 architecture with the GNU C/C++ compiler
(version~4.3.4) with optimisation flag \textsf{O3}.
The serial CPU benchmarks were done with HAL's MD package \cite[commit a628797]{Colberg2009};
it was compiled with single floating-point precision for x86\_64 using
the GNU compiler again with optimisation flag \textsf{O3},
which implies automatic vectorisation of loops.
On the x86\_64 architecture, GCC defaults to SSE floating-point arithmetic,
which ensures that throughout a calculation, a floating-point value is stored
with the precision mandated by its data type.

The compute times per MD step obtained on the GPU and the host are compared in
Figs.~\ref{fig:performance}a and~b, respectively.
They are proportional to the number of
particles $N$, and the double-logarithmic representation nicely corroborates
the linear complexity of the simulation algorithm.
In the GPU case, the linear scaling is only obeyed for sufficiently large
$N\gtrsim 20{,}000$; for smaller $N$, the compute times show a constant offset
of $0.2-0.5$\,ms, reflecting the overhead of parallel sorting and reduction.
The prefactor increases for denser systems, which we attribute to larger
neighbour lists.  At $\rho^*=0.4$, we measure a GPU time per MD step per
particle of \SI{24}{ns}; this value doubles to \SI{53}{ns} for the highest
density, $\rho^*=1.2$.
For comparison, the timings reported in Refs.~\citealp{Anderson2008} and
\citealp{Meel2008} at $\rho^*=0.4$ are \SI{62}{ns} and \SI{200}{ns}
per MD step and particle using the older NVIDIA GeForce 8800 GTX hardware with
only 128 CUDA cores at \SI{1.5}{GHz}.

For the described double-single precision implementation, the execution times
on the GPU increase modestly by about 20\%.
The dependence on system size and density is similar to the results for the
single precision implementation in \fig{performance}a.
Thus, the performance penalty for using single-double precision in critical
parts of the algorithm results roughly in a global factor.
A significant part of the additional execution time seems to be related to
memory latency.
Using double-single precision, twice the amount of data is read and written for
velocity and position vectors.
For the three-dimensional velocity vectors, this totals to $3\times 2=6$ 32-bit
words per particle, requiring at least two coalesced memory operations by using,
e.g., two arrays of \textsf{float4}.
In contrast, the velocity vector can be accessed with a single coalesced
operation for single precision.
We expect that these considerations hold also if native double precision on the
GPU is used, dropping the additional floating-point operations for the
double-single arithmetic.
CUDA devices of compute capability~2.0 only support texture read operations of up
to 128 bytes, which would entail splitting position and velocity vectors into
properly aligned \textsf{double2} and \textsf{double} components, and thus
require twice the amount of memory accesses per vector.

On the host processor, the MD step time of the serially executed programme is
proportional to $N$ for small system sizes too (\fig{performance}b).
With increasing system size, the execution times increase slightly, probably
due to a growing number of cache misses.
In particular, performance degraded substantially (by a factor of 2) for systems
of more than \num{e5} particles if the particles were randomly distributed in
memory.
Thereby, the obtained speedups were spoiled considerably, which we have solved
by regularly sorting the particles in memory as in the GPU implementation.
The prefactor spreads by about 60\% around its average, which is somewhat larger
than on the GPU (40\%).
At $\rho^*=0.4$, we have measured an MD step time per particle of
1.8\,\textmu{}s, which is comparable to the timings obtained below with the LAMMPS
package on a single processor core.

The relative performance gain of the GPU over the CPU is displayed in
\fig{performance}c.
It depends to some extend on the particle density such that dense systems are
favoured by the GPU. While the speedup factor is as small as 4 to 12 for small
systems of just \num{e3} particles, it goes up to values around 40 for \num{e4}
particles, and it reaches an approximate plateau between 70 and 80 for more than
\num{e5} particles, being close to the theoretical limit for a compute-bound
algorithm.
The LAMMPS benchmarks were performed at the Leibnizrechenzentrum in Munich
on 8-way nodes containing 4 dual-core processors of type AMD Opteron 8218 HE
\SI{2.6}{GHz}.
The tested version of the package was released on 9~January 2009, it was compiled
with Intel's C++ compiler (version 10.1) using the optimisation flags
\textsf{O2}, \textsf{ipo}, \textsf{unroll}, and \textsf{fstrict-aliasing}.
The programme was run in parallel mode using the MPI interface, the nodes
were connected via 10\,Gigabit Ethernet, and the MPI library used was from Parastation.
We have tested configurations with 8, 16, 24, 32, 48, and 64 processes.
For each particle number and
density, the fastest configuration was selected, and these smallest MD step
times are shown in \fig{performance}d; only configurations with 1, 2, 6, and 8
nodes are relevant for particle numbers between \num{e3} and \num{e6}.
For system sizes $10^3 \lesssim N \lesssim 10^4$, a single node (8 processes)
is favourable. Although the parallelisation overhead is significantly smaller than for
the GPU, the measured times per MD step exceed those on the GPU for more than
4,000 particles.
Restricting the configuration to a single 8-way node even for larger systems,
the execution times for \num{e5} particles are found to be 5 to 10 times slower
than for the GPU, depending on density; for higher particle numbers,
the single-node performance goes drastically down, probably due to frequent
cache misses.
For a large system of $10^4 \lesssim N \lesssim 10^5$ particles, the fastest
execution is obtained with 48 processes, but the time per MD step and particle
is about 2 times slower compared to the GPU, in particular at high densities.
For huge systems, $N\gtrsim 10^5$, the absolute numbers and the spread of the
execution times are comparable for the GPU and 64 processes.
In conclusion, a single GPU outperforms a conventional cluster by a factor of 2
for large systems and performs similarly as 8 recent 8-way nodes for huge system
sizes.

\section{Numerical long-time stability}
\label{sec:stability}

We have thoroughly tested the numerical long-time stability of our
implementation with respect to momentum and energy conservation.
In Section~\ref{sec:glassy}, we will show how a drift in these quantities may
affect the dynamics of the simulated system.
Numerical errors will be larger if only single precision arithmetic is used, but
the pace at which numerical errors accumulate also depends on the form of the
potential.
We will compare the Lennard--Jones potential and its purely repulsive part, also
known as Weeks-Chandler-Andersen (WCA) potential~\cite{Weeks1971},
$V_\text{WCA}(r;\epsilon,\sigma)=V_\text{LJ}(r;\epsilon,\sigma)+\epsilon$
which is cutoff at $r_c=2^{1/6}\sigma$.
This potential allows for smaller neighbour lists with less particles, speeding
up the simulation by up to 40\%.
Further, we will discuss smoothed versions of the potentials, which possess a
continuous second derivative by multiplying with the function
$g\boldsymbol(x=(r-r_c)/h \sigma\boldsymbol)=x^4/(1+x^4)$.
The following scrutiny is based on a system of $N=10{,}000$ particles at
$\rho^*=0.75$ and $T^*=1.12$.
For either the LJ or the WCA potential, all results share the same initial
configuration, which was obtained by equilibrating an initial fcc lattice in the
NVT ensemble with $\mu^*=1$, $\delta t^*=0.001$ and $h=0.005$ over \num{e5} steps
using the GPU (double-single precision) implementation.

\subsection{Momentum conservation}

\begin{figure*}
  \centering{\includegraphics{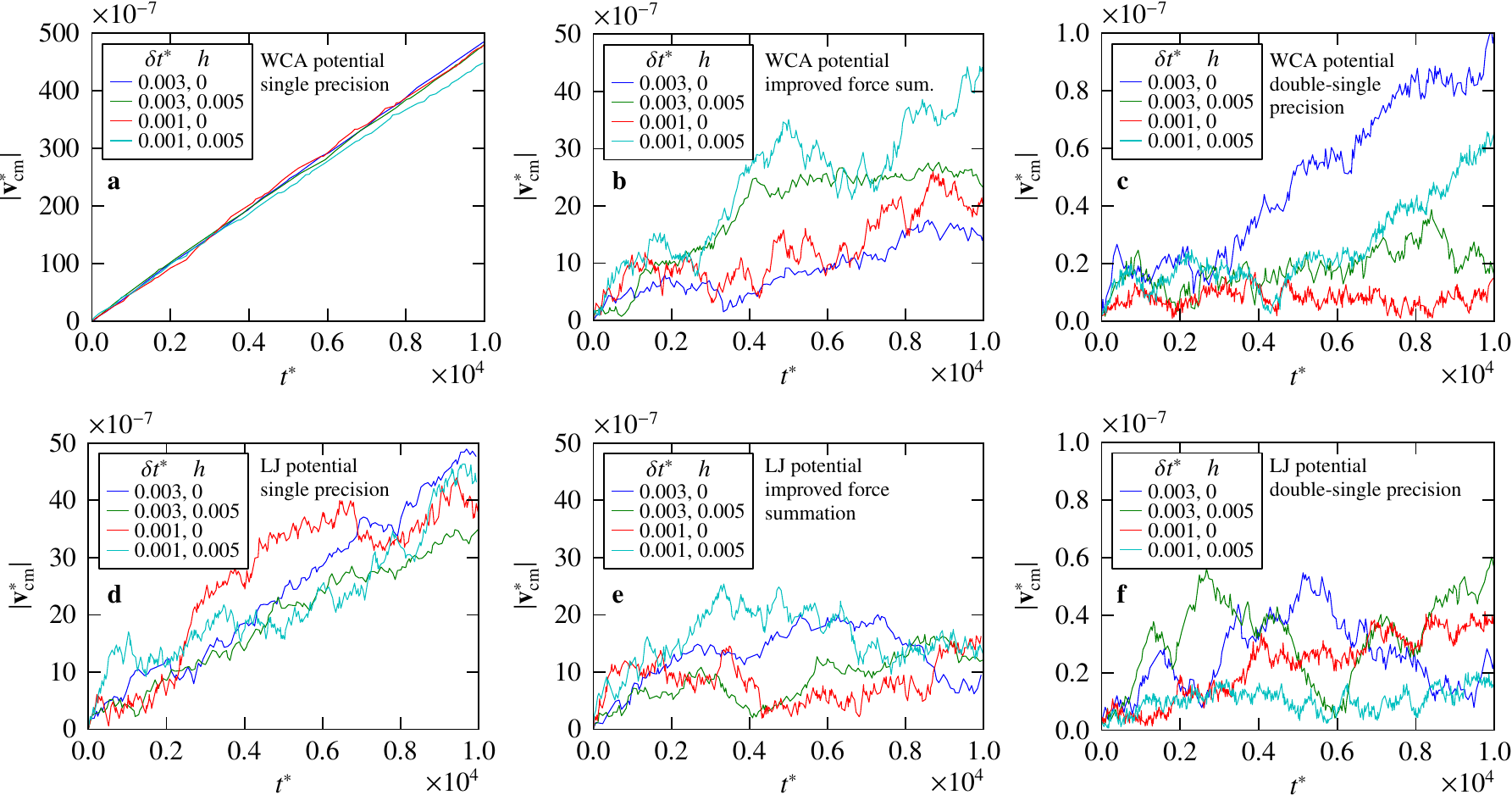}}
  \caption{Drift of the centre-of-mass velocity for simulation runs on the GPU
  if forces are accumulated in single precision and \naive{} order (panels a,d),
  or in double-single precision and opposite cell order~(b,e),
  and if double-single precision is used for both summation of forces
  and the velocity-Verlet integration~(c,f).
  The two rows compare the purely repulsive WCA potential~(a--c) and
  the Lennard--Jones potential~(d--f). Each panel contains data for two different
  time steps ($\delta t^*=0.003$ and $0.001$) with and without smoothing the
  potentials ($h=0.005$ and $0$).}
  \label{fig:momentum_conservation}
\end{figure*}

Momentum conservation implies that the interaction between particles does not
generate additional momentum or, equivalently, a drift of the centre-of-mass (c.m.)
velocity or mean particle velocity, $\vec v_\text{cm}=\expect{\vec v_i}_N$.
For the summation of forces, a standard index loop over the neighbouring cells yields first
a large force contribution to one direction, which is later cancelled by the forces from
particles on the opposite side. Such an implementation in single precision
results in a clear drift of the c.m.\ velocity for the WCA potential as
illustrated in \fig{momentum_conservation}a;
the drift is an order of magnitude smaller in the case of the LJ potential,
\fig{momentum_conservation}d.
Using double-single precision for the summation of forces and summing opposite
cells together reduces the drift by factors of 20 and 2.7 for the WCA and LJ
potentials, respectively (panels b and~e).
The drift of the c.m.\ velocity is suppressed by another factor of 100(!) in
both cases if the velocity-Verlet integration is done in double-single precision
too (panels c and~f).
Modifying the time step of the integration or smoothing the potential does not have
a significant effect on the velocity drift.

\subsection{Energy conservation}

Faithful simulations in the microcanonical ensemble crucially depend on the
conservation of the total system energy, $E=E_\text{kin}+E_\text{pot}=\textit{const}$.
This condition is particularly sensitive to limited floating-point precision
and requires particularly careful examination.
For example, an energy drift of 2\% was found after \num{e6} MD steps using the
single precision version of GROMACS (which is the default)~\cite{Lippert2007}.

\begin{figure*}
  \centering{\includegraphics{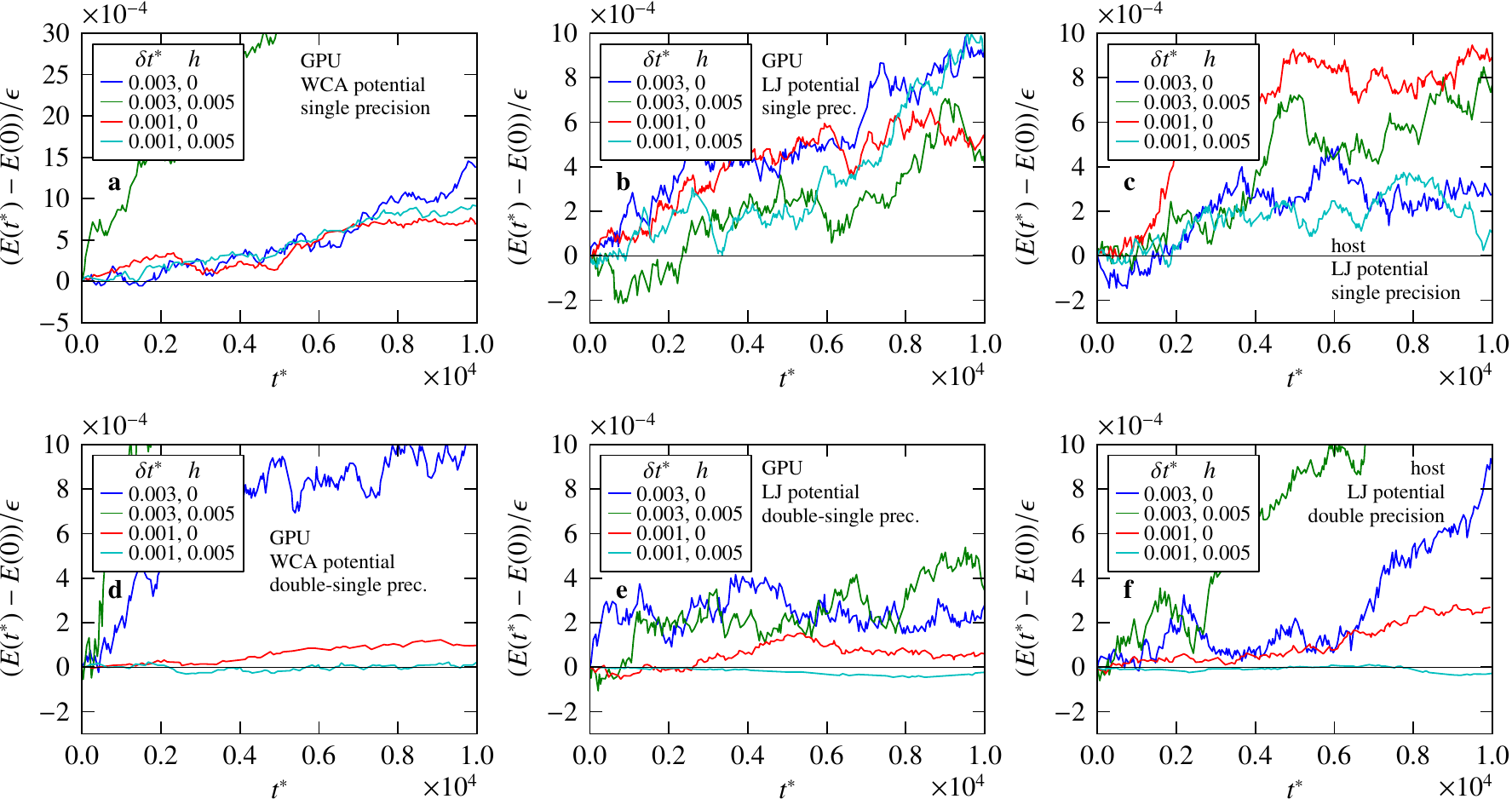}}
  \caption{Conservation of total energy per particle during a long simulation run.
  The first two columns refer to GPU results (panels a,b,d,e) and
  the last column to the host implementation~(c,f).
  Single floating-point precision was used for the top row and double-single or double
  precision for the bottom row. The sensitivity on the precision is compared
  for the WCA and LJ potentials with and without smoothing and for two different
  time steps as indicated. Note the different y-axis of panel~a.
  }
  \label{fig:energy_conservation}
\end{figure*}

\begin{figure*}
  \centering{\includegraphics{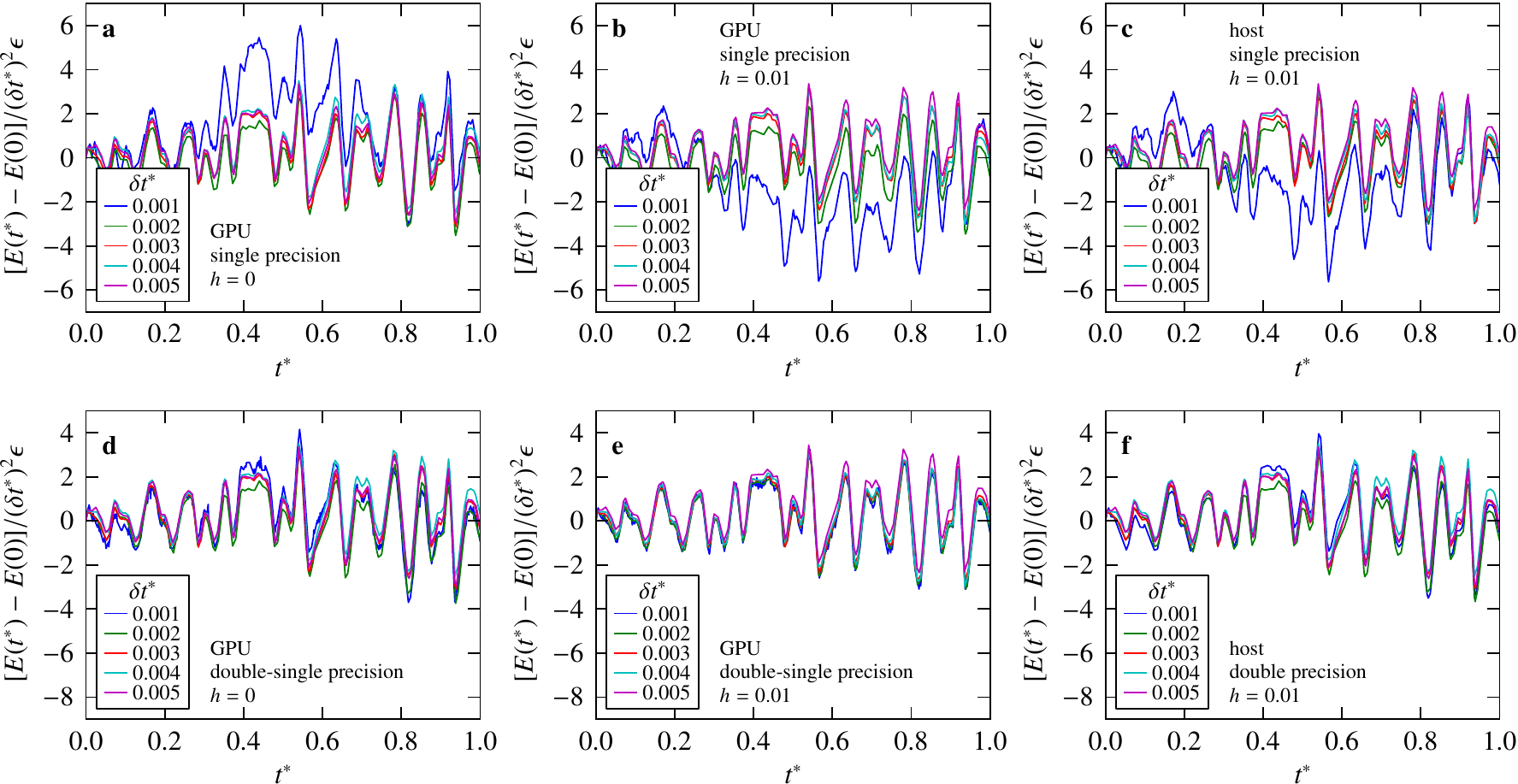}}
  \caption{Scaling test for the numerical fluctuations of the total energy around
  the initial value on the GPU (left and centre column) and the host (right column).
  Results using single precision are shown in the top row, double-single (or native double)
  precision was used in the bottom row. A smoothed LJ potential was
  used in the centre and right columns ($h=0.01$).}
  \label{fig:energy_fluctuations}
\end{figure*}

For our GPU and host implementations using both single and double precision, we
have examined the evolution of the total system energy, displayed in
\fig{energy_conservation}.
In addition, we compare the degree of energy conservation for WCA and LJ
potentials with and without smoothing and for two different time steps.
The single precision GPU implementation shows a clear drift of about 0.1\% per
\num{e7} MD steps for both potentials (panels a and~b).
There is a contribution from the c.m. velocity drift, which is, however, orders
of magnitude smaller.  Smoothing the potentials at the cutoff does not improve
energy conservation here.
For the WCA potential, the time step $\delta t^*=0.003$ is obviously too large;
energy conservation is considerably more degraded if the potential is smoothed.
Using double-single precision for the summation of forces and in the
velocity-Verlet algorithm (panels d and~e) reduces the energy drift by factors
of about 3 (LJ potential) to 7 (WCA).
Additional smoothing of the potentials, however, improves energy conservation
significantly by an extra order of magnitude.
Thereby, the drift is almost eliminated in the case of $\delta t^*=0.001$ down
to \num{5e-6} and \num{6e-5} per \num{e7} steps for the WCA and the LJ
potential, respectively.
Such a tiny drift, however, is dominated and obscured by numerical fluctuations.

The host simulation results shown in panels c and~f undermine that the
degradation of energy conservation due to single precision is not specific to
the GPU implementation.
While the quantitative evolution of the energy in panels e and~f differs due to
the implementation of floating-point arithmetic in either 44-bit double-single
precision (GPU) or native 53-bit double precision (CPU), the orders of magnitude
of energy conservation are comparable.

\subsection{Numerical energy fluctuations}

The analysis of the numerical energy fluctuations yields a sensitive test of the
numerical accuracy which requires only short simulation runs.
Newton's differential equations are discretised by the velocity-Verlet algorithm
to first order in the time step $\delta t$, introducing a discretisation error of
order $\delta t^2$---provided the potential possesses a continuous second
derivative.
Thus, one expects that the total energy shows numerical fluctuations around
its initial value which scale as $\delta t^2$.
In particular, $[E(t)-E(0)]/\delta t^2$ should roughly be independent of the
time step.

For time steps between $\delta t^*=0.001$ and 0.005, these rescaled energy
fluctuations are displayed in \fig{energy_fluctuations} for the LJ potential.
Using double-single precision on the GPU, all five curves nicely collapse
(panel~d) over the range of 200 to 1000 integration steps.
Merely the largest time step, $\delta t^*=0.005$, is slightly off, indicating
that higher order terms become relevant in this case.
Smoothing the potential at the cutoff improves the collapse, from which
small time steps benefit especially (panel~e).
If only single precision arithmetic is used on the GPU, the collapse is poor
and smoothing the potential at the cutoff has essentially no effect (panels a and~b).
The two smallest time steps are completely off, which we
attribute to the quick accumulation of rounding errors from the tiny increments.

For comparison, we have added the results from the host implementation in
single and double precision (panels c and~f).
The almost identical behaviour as on the GPU corroborates that the discussed
effects are due to the limited precision only, and no other numerical artifacts
are introduced by the GPU.
An insignificant, but nevertheless interesting difference between the GPU
(double-single precision) and the CPU (double precision) results are the
high-frequency fluctuations visible at the smallest time step $\delta t^*=0.001$
for the GPU case (panels d and~e).
These tiny fluctuations are caused by the evaluation of the individual pair
force contributions in single precision.
We verified that the high-frequency fluctuations disappear if the entire force
algorithm on the GPU is implemented in double-single precision at the cost of
significantly reduced performance.

\begin{figure*}
  \centering{\includegraphics{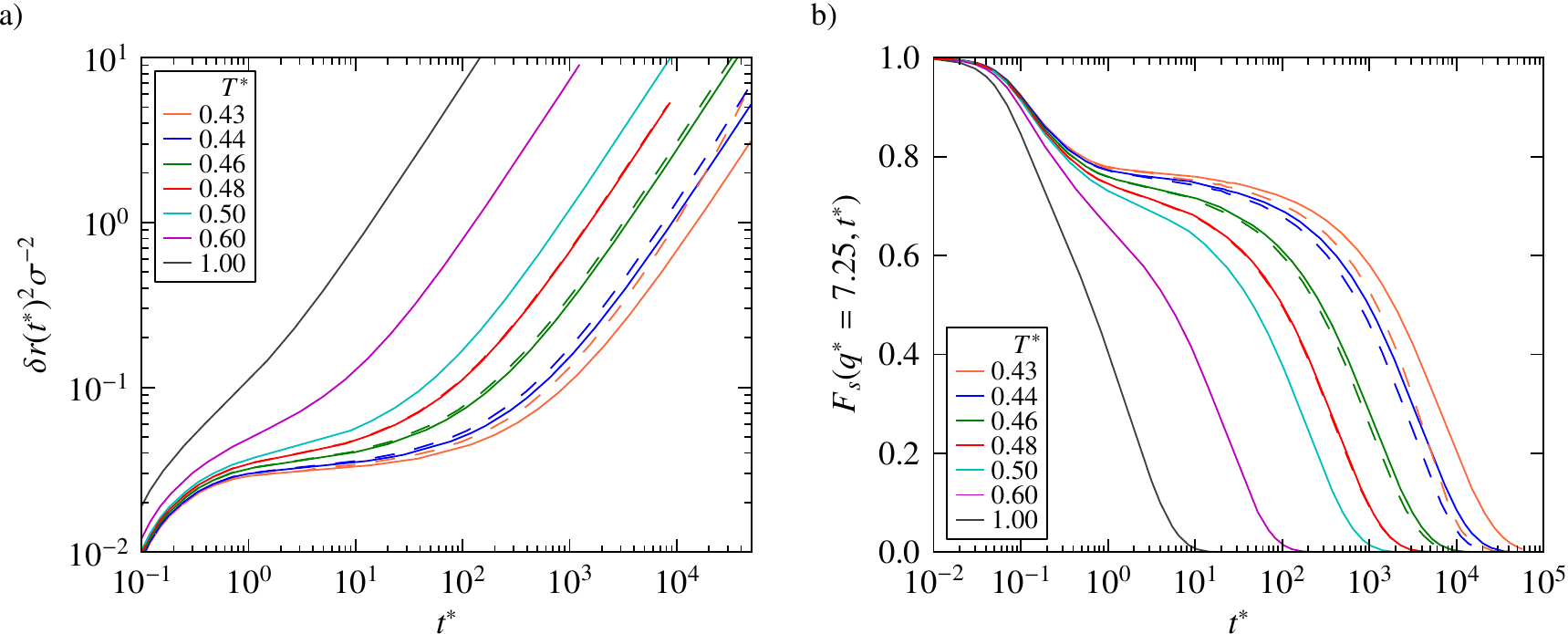}}
  \caption{Simulation results for a Kob--Andersen mixture obtained on a GPU using
  double-single arithmetic.
  Mean-square displacements (a) and incoherent intermediate scattering functions (b)
  of the A particles are shown for different temperatures approaching the glass
  transition at fixed density $\rho^*=1.2$.
  Broken lines indicate results with single floating-point precision.}
  \label{fig:msd+isf}
\end{figure*}

\section{Application to glassy dynamics of a supercooled liquid}
\label{sec:glassy}

We employed the above GPU implementation of a molecular
dynamics simulation to reproduce the slow dynamics of a supercooled liquid of soft
spheres.
We resorted to the Kob--Andersen (KA) binary mixture~\cite{Kob1994, Kob1995,
Kob1995b}, which has proven
to be a useful model system that reliably delays crystallisation.%
\footnote{We could not find any signs of crystallisation for a very long
thermostat run of a small system of 1,500 particles at $T^*=0.4$ and
$\rho^*=1.2$ over a time span of \SI{e8} LJ units or \SI{5e9} MD steps.}
Specifically, we have investigated a relatively large system of 40,000~A and
10,000~B particles of equal masses $m$ interacting with Lennard--Jones
potentials,
$V_{\alpha\beta}(r)=V_\text{LJ}(r;\epsilon_{\alpha\beta},\sigma_{\alpha\beta})$
with the parameters chosen as in Ref.~\citealp{Kob1994}:
$\sigma_{AA}=1$, $\sigma_{BB}=0.88$, $\sigma_{AB}=0.8$,
$\epsilon_{AA}=1$, $\epsilon_{BB}=0.5$, and $\epsilon_{AB}=1.5$.
The time step of the Verlet integrator was taken to be
$\delta t^*=0.001$ in units of $\sqrt{m\sigma_{AA}^2 /\epsilon_{AA}}$.
Equilibration runs were done for fixed energy and covered about 10~times the
structural relaxation time; all results were obtained for two independent
systems with different initial conditions.
Production runs of \num{1e6} to \num{2e8} steps (for $T^*=0.60$ and $0.43$,
respectively) including the online evaluation of the correlation functions
merely took between 1.4 hours and 9.6 days of real time.

For the following discussion, we need some basic quantities central to the
description of glassy dynamics.
The simplest quantity to characterise the transport dynamics is the mean-square
displacement (MSD) $\delta r^2(t) = \dexpect{\vec r_i(t)-\vec r_i(0)}_N$, where
$\expect{\cdot}$ denotes a microcanonical time average and $\expect{\cdot}_N$ an
average over all particles, $i=1,\dots, N$.
Here, we restrict the discussion to A~particles.
The diffusion constant $D$ of a tagged particle is then obtained via
$\delta r^2(t)\simeq 6 D t$ for $t\to\infty$.
A more complex quantity is the time-correlation function of local density
fluctuations, the intermediate scattering function (ISF).
The self-part of the ISF is defined as
$F_s(\vec q,t)=\dexpect{\rho^{(s)}_{\vec q}(t)\rho^{(s)}_{-\vec q}(0)}_N$
using the Fourier components of the tagged particle density,
$\rho^{(s)}_{\vec q}(t)=\exp\left(-\mathrm{i}\vec{q}\dotprod\vec r^{(s)}(t)\right)$,
with discrete wavenumbers $\vec q\in (2\pi/L)\mathbb{Z}^3$
and $\vec r^{(s)}$ denoting any of the $\vec r_i$.
Rotational symmetry implies that $F_s(\vec q,t)=F_s(q,t)$ merely depends on the
magnitude of the wavevector $q=|\vec q|$,
and additional averaging can be done over the orientation of~$\vec q$.
To quantify the structural relaxation time $\tau_\alpha$, we adopt the usual definition
$F_s(q_\text{max},\tau_\alpha)=1/\text{e}$, where $q_\text{max}=7.25 \sigma^{-1}$
denotes the maximum of the structure factor.

The MSD develops a pronounced plateau for decreasing temperature,
reflecting the caging by the arrested surroundings; see \fig{msd+isf}a.
The diffusion coefficient is drastically suppressed as the temperature approaches
an anticipated glass transition temperature $T_g$
and spans more than 3~decades over the simulated temperature range.
Correspondingly, the density correlators shown in \fig{msd+isf}b
develop a plateau, which is a signature of the  structural arrest.
The slow dynamics is quantified by the diverging structural relaxation time
$\tau_\alpha$, for which we observe an increase by 4 orders of magnitude.

\begin{table}
  \centering\small \tabcolsep=3pt
  \begin{tabular}{c|ccc|ccc}
    \hlx{hhv}
    $T^*$ & $D^*_\text{d.-s.}$ & $D^*_\text{single}$ & error &
    $\tau^*_{\alpha;\,\text{d.-s.}}$ & $\tau^*_{\alpha;\,\text{single}}$ & error \\
    \hlx{vhv}
     0.48 & \num{1.02E-4}  & \num{1.03E-4} & \hphantom{0}1\% & \num{2.31E3} & \num{2.26E3} & \hphantom{0}2\% \\
     0.46 & \num{4.48E-5}  & \num{5.12E-5}  & 14\% & \num{6.43E3} & \num{5.69E3} & 12\% \\
     0.44 & \num{1.73E-5}  & \num{2.43E-5}  & 40\% & \num{2.19E4} & \num{1.71E4} & 22\% \\
     0.43 & \num{1.02E-5}  & $\infty$ &  & \num{4.24E4} & \num{2.50E4} & 41\% \\
    \hlx{vhh}
  \end{tabular}
  \caption{Dimensionless diffusion constants $D^*$ and relaxation times $\tau^*$
  of A particles for temperatures close to the glass transition.
  The table compares simulation results obtained with double-single and single
  floating-point precision and gives the relative error due to the limited
  precision.}
  \label{tab:diffusion+tau_alpha}
\end{table}

A clear observation of the glassy dynamics requires sufficient separation between
short-time features and the slow structural relaxation.
Hence, the temperature has to be fine-tuned close to the transition,
presupposing a sharp value for $T_g$.
Moreover, simulation runs become very long and the energy of the system
must be extremely stable during a complete run;
we could limit the energy drift to merely \num{3e-5} over \num{2e8} MD steps at $T^*=0.43$.
It turns out that such a long-time stability cannot be maintained with single
floating-point precision.
For low temperatures $T^*=0.43, 0.44, 0.46, 0.48$, we have performed additional
production runs with single precision, and both the MSD and the density
correlators deviate significantly at long times, see \fig{msd+isf} (broken lines).
The system heats up during the simulation, which introduces physical artifacts
in the form of a faster relaxation at the pretended temperature.
The determined diffusion constants and structural relaxation times
for both levels of precision are compared in Table~\ref{tab:diffusion+tau_alpha},
revealing quantitative differences of up to 41\%.
At the lowest investigated temperature, the system appears to become even
super-diffusive at long times; note the crossing of the MSD curve at $T^*=0.43$
obtained with single precision and the one at $T^*=0.44$ using double-single
precision in \fig{msd+isf}a.

Further, the use of current graphics processors facilitates the investigation
of much larger system sizes than were usually accessible before.
Among other benefits, data quality is enhanced as statistical fluctuations of
tagged particle quantities are expected to scale as $1/\sqrt{N}$.
First results for the velocity autocorrelation function display an
excellent signal-to-noise ratio and shed light on novel power-law correlations
at low temperatures~\cite{Glassy_VACF:2010}.

\section{Conclusion}
\label{sec:conclusion}

We have shown how recent graphics processing units can be harnessed to
carry out large-scale molecular dynamics simulations in the microcanonical ensemble
with strict energy conservation even after \num{e8} MD steps.
Using GPU computing, we were able to reproduce the slow glassy dynamics of a
binary Lennard--Jones mixture over 4~nontrivial decades in time.
Single floating-point precision, however, is not sufficient for this purpose and
may result in qualitatively and quantitatively wrong results;
e.g., the diffusion coefficient was found to diverge at $T^*=0.43$ and to deviate
by up to 40\% at higher temperatures due to the limited energy conservation.
We have shown that the mediocre native double precision performance of recent
GPUs can be overcome by implementing numerically critical parts of the MD
algorithm with double-single precision floating-point arithmetic, which is based
on single precision instructions.

The described MD simulation package is fully implemented on the GPU using CUDA,
and it avoids costly memory transfers of trajectory data between host and GPU.
In addition to earlier work~\cite{Meel2008, Anderson2008}, the sorting of
particles in memory and the evaluation of dynamic correlation functions are
completely performed on the GPU.
The number of simulated particles is solely limited by the amount of global
device memory.
On the NVIDIA GeForce GTX 280 providing \SI{1}{GB} of memory, simulations of
864,000 particles are possible, but the barrier of one million particles is
broken on the NVIDIA Tesla C1060 with \SI{4}{GB} of (somewhat slower) memory.
Our performance measurements show speedups of 70 to 80 compared to a serial
simulation on the host processor, and we have shown that the GPUs we have
deployed perform similarly to LAMMPS on a modern, conventional HPC system
running in parallel on 64 processor cores.
We have found that the use of double-single precision in specific parts of the
algorithm increases the execution times by merely 20\%, which we attribute mainly
to a doubling of memory accesses.
While the use of native double precision arithmetic in the next generation of
GPUs will reduce the number of floating-point operations compared to
double-single precision, we expect that the performance penalty due to the
latency of global memory access will remain.
In particular, the trade-off between performance optimisation and numerical
accuracy in terms of floating-point precision will persist for GPUs as it does
for conventional processors.

In summary, current graphics processors provide a powerful and robust means
for state-of-the-art simulations of simple and complex liquids in general and
for numerical studies on glass-forming liquids in particular.
Substantial computing resources can be delivered already by local GPU clusters
containing a few dozen high-end GPUs, which are affordable in terms of
acquisition cost and maintenance for a single institute and which are likely to
play a considerable role in future simulation-based research.
In addition, some national computing centres have started to support
GPU-accelerated computing on large dedicated GPU clusters, which are useful at
the single-GPU level already. They will, however, be fully exploited only
by the further development of simulation packages running on distributed GPUs
(see Refs.~\citenum{Xu:2010, Voelz:2010, Harvey:2009, Block2010} for examples),
enabling the routine investigation of large and complex systems
that can be studied today on exceptionally few supercomputers only.

\section{Acknowledgment}

We thank Thomas Franosch and J\"urgen Horbach for many helpful discussions and
Erwin Frey for generous support.
F.H.\ acknowledges financial support from the Nanosystems Initiative Munich (NIM).
The described software, \emph{HAL's MD package}, is licensed under the GNU General
Public License and is freely available at the project's website~\cite{Colberg2009}.

\bibliographystyle{elsarticle-num}
\providecommand{\urlprefix}{} 
\bibliography{glassy_gpu}

\end{document}